# The Impact of Speed and Bias on the Cognitive Processes of Experts and Novices in Medical Image Decision-making

Jennifer S. Trueblood[a,1,2] (jennifer.s.trueblood@vanderbilt),
William R. Holmes[b,1,2] (william.holmes@vanderbilt.edu),
Adam C. Seegmiller[c] (adam.seegmiller@vanderbilt.edu),
Jonathan Douds[c] (jonathan.j.douds@vanderbilt.edu),
Margaret Compton[c] (margaret.l.compton@vanderbilt.edu),
Eszter Szentirmai[c] (eszter.szentirmai@gmail.com)
Megan Woodruff[a] (megan.e.woodruff@vanderbilt.edu),
Wenrui Huang[a] (wenrui.huang@vanderbilt.edu),
Charles Stratton[c] (charles.stratton@vanderbilt.edu),
Quentin Eichbaum[c,1] (quentin.eichbaum@vanderbilt.edu)

[a]Department of Psychology, Vanderbilt University
[b]Department of Physics and Astronomy, Vanderbilt University
[c]Vanderbilt Pathology Education Research Group (VPERG), Department of Pathology, Microbiology and Immunology, Vanderbilt University Medical Center (VUMC).

**Abstract**

Training individuals to make accurate decisions from medical images is a critical component of education in diagnostic pathology. We describe a joint experimental and computational modeling approach to examine the similarities and differences in the cognitive processes of novice participants and experienced participants (pathology residents and pathology faculty) in cancer cell image identification. For this study we collected a bank of hundreds of digital images that were identified by cell type and classified by difficulty by a panel of expert hematopathologists. The key manipulations in our study included examining the speed-accuracy tradeoff as well as the impact of prior expectations on decisions. In addition, our study examined individual differences in decision-making by comparing task performance to domain general visual ability (as measured using the Novel Object Memory Test (NOMT) (Richler et al., 2017). Using Signal Detection Theory (SDT) and the Diffusion Decision Model (DDM), we found many similarities between expert and novices in our task. While experts tended to have better discriminability, the two groups responded similarly to time pressure (i.e., reduced caution under speed instructions in the DDM) and to the introduction of a probabilistic cue (i.e., increased response bias in the DDM). These results have important implications for training in this area as well as using novice participants in research on medical image perception and decision-making.

# Significance Statement

The ability to classify and interpret medical images is critical in the diagnosis of many diseases. Despite significant improvements in imaging assays as well as meticulous education and training, diagnostic errors still occur. In order to improve diagnostic decision-making based on medical images, it is critical to understand the cognitive processes involved in these decisions. This research borrows well-validated experimental and computational methods from perceptual decision-making and applies them to investigate cancer cell image identification. Using both non-experts as well as pathologists (residents and faculty), we examine the impact of time pressure and externally imposed bias on the identification of single cell images related to cancer diagnosis. Using computational modeling techniques, we find that these manipulations have important impacts on diagnostic decisions. Specifically, we find similarities in how novices and pathologists tradeoff speed and accuracy instructions as well as how they respond to externally imposed bias. In addition, we find that participants with better domain general visual ability perform better at the task. In sum, these results shed light on the cognitive mechanisms that play a role in medical image perception and decision-making. In the future, this knowledge could be used to improve training and education, and this method of investigation could lead to new insights about the cognitive processes involved in image-based decisions.

Introduction

Accurate interpretation and classification of medical images is an important component of the diagnosis and treatment of numerous diseases. A wide range of medical disciplines (Samei & Krupinski, 2010) ranging from pathology (our focus here), to radiology, to ophthalmology rely on expert analysis of images to detect abnormalities. While the exact rate of diagnostic errors is unknown, consistent evidence suggests error rates are >10% (Goldman et al., 1983; Hoff, 2013; Kirch & Schafii, 1996; Shojania, Burton, McDonald, & Goldman, 2003; Sonderegger-Iseli, Burger, Muntwyler, & Salomon, 2000). It is thus critical that we understand how people make perceptual decisions from medical images in order to improve training and minimize the occurrence of misdiagnoses. This requires investigation of the cognitive processes underlying decision-making in this domain and how those processes evolve with training and experience. The goal of the present paper is to use experimental methods and computational tools developed in the area of perceptual decision-making to probe the cognitive processes involved in pathology image based decisions in novices and experts.

Decisions based on medical images have a number of parallels with perceptual decision-making, where people make choices based on sensory information. The investigation of perceptual decision-making has a rich history in psychology, cognitive science, and neuroscience. In aggregate, this research has shown that perceptual decisions are typically based on the accumulation of information over time. Such accumulated perceptual information is thought to be related to neural activity in multiple cortical and subcortical brain areas (Gold & Shadlen, 2007; Heekeren, Marrett, & Ungerleider, 2008; Summerfield & de Lange, 2014; Summerfield & Egner, 2009). This accumulation process is known to be influenced by external factors such as time pressure and expectations (Egner, Monti, & Summerfield, 2010; Leite & Ratcliff, 2011; Maddox & Bohil, 1998; Mulder, Wagenmakers, Ratcliff, Boekel, & Forstmann, 2012). Computational modeling has shown that these different external factors influence different latent components of the decision process. In particular, time pressure affects response caution (quantified by the amount of information needed to make a decision) while prior expectations impact internal biases (e.g., bias towards reporting the presence of an abnormality even before viewing an image) (Leite & Ratcliff, 2011; Mulder et al., 2012).

However, perceptual decision-making of medical images in clinical settings has received less attention. Numerous studies have probed the perceptual processes involved in image based decisions, particularly in the context of radiology (Bertram, Helle, Kaakinen, & Svedstrom, 2013; Krupinski, 2010; van der Gijp et al., 2017). However, these studies have largely focused on how medical image observers perform visual search (Bertram et al., 2013; Krupinski, 2010; Krupinski, Graham, & Weinstein, 2013; Krupinski et al., 2006; van der Gijp et al., 2017). Eventually however, a decision must be made and understanding the cognitive processes involved in these decisions is the main objective of this paper.

Here, we present a study investigating the cognitive processes underlying cancer image detection in diagnostic pathology. More specifically, we investigate how various external factors influence the ability of novice undergraduate students and pathologists (residents and faculty) to distinguish between normal (standard white blood cells such as lymphocytes, monocytes, or neutrophils) and abnormal cancer cells ("blast" cells, associated with acute leukemia) in clinical images. Toward this end, we take a joint experimental and modeling approach utilizing experimental paradigms and modeling methods previously developed in the course of basic research on perceptual decision-making (Ratcliff & Smith, 2004; Schouten & Bekker, 1967; Wickelgren, 1977).

To investigate this process experimentally, we passively collected a large bank of digital images of both blast and non-blast white blood cells drawn from patients at the Vanderbilt University Medical Center (all images were obtained as part of routine clinical care). A panel of expert pathologists classified each of these images, providing a fully annotated data set consisting of hundreds of images of varying type and level of difficulty. Using this image bank, we developed a perceptual decision-making experiment to investigate how time pressure and externally imposed bias influence individuals' behavior.

We chose to examine the speed-accuracy tradeoff (SAT) (Reed, 1973; Wickelgren, 1977) as well as the impact of external bias because these factors have relevance in the clinical context. With the current and projected shortage of medical technologists and pathologists (Allen, 2002; Bennett, 2015; Lewin, 2016; Sullivan, 2016) coupled with a desire to improve throughput and turnaround times and reduce costs, many laboratories hope to increase productivity by using automated basic recognition sorters. In essence, automated systems have the potential to offset some of the human workload in order to increase productivity, which is largely dependent on the speed with which slides are screened. For example, the FDA increased the workload for cytotechnologists from 100 slides per day to 200 slides per day if they are using the ThinPrep imaging system, an automated system used for gynecologic cytology (Elsheikh et al., 2013). However, it is unclear how this increase in workload (even though it comes with the assistance of an automated system) influences diagnostic decisions. In particular, research has shown that decreasing screening times for cytotechnologists from 5 minutes per slide to 3.7 minutes per slide resulted in a lower detection of abnormal findings (10.4% to 8.3%) and an increase in false negatives (3.8% to 7.0%) (Elsheikh et al., 2010). In other words, the cytotechnologists were trading off speed and accuracy, even when they had access to an automated system. More generally, as machine learning and AI become more integrated into the diagnostic process, the desire for increased productivity is likely to result in higher workloads for medical image observers. While machines will likely be able to process images faster than humans, human observers will still need to be part of the diagnostic process (at least for the foreseeable future). Thus, it is critical to understand how medical image observers tradeoff speed and accuracy in diagnostic decision-making.

In addition, prior expectations and biases are likely to play a significant role in medical image-based decision-making. In diagnostic pathology, images may be passed through automated basic recognition sorters and/or analyzed by medical technologists and residents before being analyzed by senior faculty experts. In this diagnostic chain, images that clearly lack abnormalities are rarely passed up the chain. Thus, an image that has made it to a senior faculty expert's desk may in and of itself be a cue, setting expectations before an image is even seen.

In addition to testing participants' ability to discriminate between and classify images of blast and non-blast cells, we also investigate how participants' domain general visual ability impacts their performance on this task. Toward this end, we employ a second task, the Novel Object Memory Test (NOMT), to assess each participant's general ability to learn and recognize objects that they have no prior experience with (Richler, Wilmer, & Gauthier, 2017). We use this to probe to what extent general object recognition, which has been studied in much more detail in lab settings (Gauthier et al., 2014; Hildebrandt, Wilhelm, Herzmann, & Sommer, 2013; McGugin, Gatenby, Gore, & Gauthier, 2012; McGugin, Richler, Herzmann, Speegle, & Gauthier, 2012), correlate with or impact participants' efficacy on the blast cell identification task.

To gain further insight into the cognitive processes underlying decisions on this task, we utilize computational modeling linked with results of this experiment. One of the benefits of quantitative modeling, and the reason we use it here, is that it provides a way to quantify latent cognitive processes and statistically separate the different components of the decision process (caution, bias, and rate of information uptake) that are not accessible through traditional statistical methods alone. For this, we utilize a version of the classic Diffusion Decision Model (DDM) (Ratcliff, 1978; Ratcliff & McKoon, 2008; Ratcliff, Smith, Brown, & McKoon, 2016), which has been shown to account for detailed patterns of behavior across a wide range of decision-making paradigms (Ratcliff, Love, Thompson, & Opfer, 2012; Ratcliff, Thapar, & McKoon, 2001, 2004, 2010), to model the choice and response time behavior of participants on this task and extract these underlying cognitive parameters.

**Experimental Methods**

*Participants*. We recruited both novice and medical professionals to complete the experiment. 37 undergraduate students at Vanderbilt University participated in exchange for course credit. In addition, 19 pathologists from the Vanderbilt University Medical

Center (VUMC) participated in exchange for a $15 gift card. We recruited pathologists with different levels of experience ranging from first year pathology residents to faculty pathologists. We targeted about equal numbers of "experienced" and "inexperienced" practitioners, defined by the number of hematopathology rotations completed. All pathology residents at VUMC must complete at least four rotations by the end of their residency. We classified individuals who completed all four mandatory rotations as "experienced" and those that had not as "inexperienced". We had 9 "experienced" and 10 "inexperienced" participants. Note that our sample sizes were based off convenience (in the case of the pathologists) as well as modeling requirements. The typical sample size for experiments using similar modeling methods is 20-40 participants (Dutilh et al., 2012; Holmes, Trueblood, & Heathcote, 2016; White & Poldrack, 2014). The data are available on the Open Science Framework at https://osf.io/r3gzs/.

*Materials*. To create the stimuli, we collected a bank of 840 digital images of Wright-stained white blood cells taken from anonymized patient peripheral blood smears at VUMC. The images were taken by a CellaVision DM96 automated digital cell morphology instrument (CellaVision AB, Lund, Sweden). This instrument and accompanying software identifies and images single white blood cells, and classifies them into one of 17 cell types based on morphologic characteristics. The classification of each cell is confirmed by a trained medical technologist.

A ratings panel of three hematopathology faculty from the Department of Pathology at VUMC was used to identify and rate each image. The raters first identified each image as a blast or a non-blast cell. Following this identification, they were asked to provide a difficulty rating for each image on a 1-5 scale. If the raters identified the image as a blast cell, they were asked to rate how similar the image was to a classic blast cell (with a rating of 1 being 'not similar' and a rating of 5 being 'very similar'). Raters were told that a classic blast cell image is one that might be used in a textbook. If raters identified the image as a non-blast cell, they were asked to rate how morphologically similar the cell is to a blast cell (with a rating of 1 being 'not similar' and a rating of 5 being 'very similar').

Out of the original set of 840 images, the three expert raters agreed on the cell type (i.e., blast or non-blast) for 633 images. From this set, we grouped the images into four types based on the difficulty ratings. The average rating for the blast images was 4.40 (SD= 0.46) and the average rating for the non-blast images was 1.68 (SD = 0.55). Blast images that received an average rating of 4.66 or greater were categorized as easy blasts (151 images, Fig 1a). Blasts images that received an average rating of 4 or less were categorized as hard blasts (98 images, Fig 1b). We selected 4.66 and 4 as the cutoff points for easy and hard images respectively because these values represented the 0.75 and 0.25 quantiles of the ratings for the blast images respectively. Images with an average rating between 4 and 4.66 were not included as we wanted clear separation between easy and hard images. Non-blast images that received an average rating of 1.66 or less were categorized as easy non-blasts (129 images, Fig 1c). Non-blasts images that received an average rating of 2 or greater were categorized as hard non-blasts (108 images, Fig 1d). We selected 1.66 and 2 as the cutoff points for easy and hard images because these values represented the 0.25 and 0.75 quantiles of the ratings for the non-blast images respectively. Images with an average rating between 1.66 and 2 were not included as we wanted clear separation between easy and hard images. For the experiment, we selected 75 images from each category for a total of 300 unique images. This bank of images was used for all aspects of the experiment including training, practice, and the main task. Note that we have more than 300 trials in the experiment, so some images are repeated (this includes repeating images from the training and practice in the main task). However, no images were repeated until all of the images from a category had been shown.

*Procedure*. In the main task, participants first completed a training stage to familiarize themselves with blast cells (both novice and experts completed the training for consistency). The training focused on teaching participants to identify blast cells and was patterned off of the training in the NOMT. There were four blocks of training trials. Each block started with participants studying five images of blast cells one at a time. After studying these five images, participants then completed 15 trials where they were presented three images (1 blast image and 2 non-blast images) and asked to choose the image they thought was the blast cell. The four training blocks had the following structure of blast and non-blast images: block 1 was easy blast versus easy non-blast, block 2 was easy blast versus hard non-blast, block 3 was hard blast versus easy non-blast, and block 4 was hard blast versus hard non-blast. Note that the image training used a total of 180 unique images (60 blast images and 120 non-blast images) out of the original set of 300.

After completing the four training blocks, participants completed a practice block of 60 trials to familiarize themselves with the main task. Each trial started with a fixation cross displayed for 250 milliseconds (ms). After fixation, participants were shown a single image and had to identify it as a blast or non-blast cell. Participants received trial-by-trial feedback about their choices in this block, thus these trials acted as additional training for the two categories of images. In this practice block, half of the trials were blast cells and half were non-blast cells. Thus participants had an equal amount of practice with each category. Across both the training and practice blocks, participants completed 120 trials (60 training and 60 practice) before starting the main task. These 120 trials used a total 150 non-blast images (corresponding to all of the non-blast images in our original set of 300 images) and 90 blast images.

The main task consisted of six blocks with 100 trials in each block. The main task was the same as the practice block where participants were ask to identify single images. However, participants did not receive trial-by-trial feedback about their choices. They received feedback about their performance at the end of each block. The 100 trials in each block were composed of equal numbers of easy blast images, hard blast images, easy non-blast images, and hard non-blast images, fully randomized.

There were three manipulations across blocks: accuracy, speed, and bias. In the accuracy blocks, participants were instructed to respond as accurately as possible and were given 5 seconds to respond. In the speed block, participants were instructed to respond quickly and were given 1 second to respond. If they responded after the deadline, they received the message "Too Slow!" The 5 s and 1 s response windows for the accuracy and speed conditions respectively were based on the response time data from the three expert raters. The 0.975 quantile of the expert raters' response times was 4.96 s, thus we set the accuracy response window to 5 s. The 0.5 quantile of the expert raters' response times was 1.04 s, thus we set the speed response window to 1 s.

In the bias blocks, participants were shown a probabilistic cue on half of the trials. The cue was a red dot that was shown after fixation for 500 ms. The cue identified the upcoming image as most likely being a blast cell. The cue was valid 65% of the time and participants were instructed about the validity at the start of the block. The validity of the cue was based on previous literature using similar cueing manipulations (Dunovan, Tremel, & Wheeler, 2014; Forstmann, Brown, Dutilh, Neumann, & Wagenmakers, 2010; Glockner & Hochman, 2011). In particular, we selected a cue with low validity because we hypothesized that a low validity cue might have a larger impact on novice participants than pathologists. That is, novices might rely more on the cue as compared to experts, who might simply ignore the cue because of its low validity. The order of the first three blocks was randomized, but with the constraint that there was one block for each type of manipulation (i.e., accuracy, speed, and bias). The order of the last three blocks was identical to the order of the first three blocks.

After completing the main task, participants completed a version of the Novel Object Memory Test (NOMT) (Richler et al., 2017). The NOMT is modeled after the Cambridge Face Memory Test (Duchaine & Nakayama, 2006) and provides a measure of domain general visual ability. In our experiment, we used two categories of novel objects (Ziggerins shown in Fig 1e and Greebles shown in Fig 1f). For each category, participants started with a learning phase where a target object was shown in three views followed by three test trials where the target was shown alongside two distractor objects. Participants received trial-by-trial feedback during these trials. This learning procedure was repeated for 6 target objects (the six targets for each category are shown in Fig 1e and 1f) where each target object was slightly different from the other targets in the same category. Following the learning phase, participants completed 54 test phase trials where they had to select which of three objects was any one of the six studied targets.

**Modeling Methods**

*Signal Detection Theory (SDT)*. We fit an equal-variance form of SDT to the data using hierarchical Bayesian methods (Lee & Wagenmakers, 2013). SDT has two main parameters of interest: discriminability and criterion. We performed separate hierarchal fits to the novice participants, inexperienced pathologists (less than 4 hematopathology rotations), and experienced pathologists (4 or more hematopathology rotations).

*Diffusion Decision Model*. To gain insight into participants decision process beyond what is possible with signal detection theory and statistical analysis of behavioral results alone, we use the canonical Diffusion Decision Model (DDM) of decision making. The DDM posits that over the course of a decision, evidence stochastically accumulates over time until a confidence threshold is reached and a decision is initiated. This model has three core elements that makes it valuable in assessing participants behavior on the blast identification task. 1) How fast people accumulate evidence over time is encoded in an *accumulation rate* parameter (*d*). A high / low rate indicates better / worse performance on the task. 2) The level of confidence a person requires to make a decision (i.e. level of caution) is encoded in a *threshold* parameter (*a*). 3) Finally, any initial preference for responding one way or the other prior to seeing the stimulus is described by a *bias* parameter (*z*). These are the three critical characteristics / parameters in the model that we will rely on to make inferences. See Figure 2 for a schematic description of the DDM.

The full version of the DDM that we use here is comprised of 9 (or 10) free parameters: accumulation rates for easy and hard blast images ($d_{BE}$, $d_{BH}$), accumulation rate for easy and hard non-blast images ($d_{NBE}$, $d_{NBH}$), trial to trial variability in those accumulation rates ($s_d$), start point ($z$, which determines the initial bias), trial to trial variability in the start point ($s_z$), evidence threshold ($a$), encoding and response time ($t_{ND}$). There is also a parameter encoding within trial stochasticity ($s$). However as is common, we fix this parameter to a value $s=0.1$ to avoid parameter degeneracy in the model (one parameter must be fixed). For the cueing instruction data, we introduce an additional parameter to denote the bias on trials where the cue is actually shown ($z_{cue}$). This will allow us to determine if the cue has any discernible effect on initial bias. Given that the speed, accuracy, and cueing instruction conditions all have the potential to influence people's behavior in different ways, we fit each instruction condition separately and do not assume up front that any model parameters are the same across experimental conditions.

We use a hierarchal Bayesian algorithm to fit DDM to the participants' data, providing an account of the choices made and the full distribution of response times at both the individual and population levels. For purposes of hierarchal DDM model fitting, we grouped all 19 pathologists (experienced and inexperienced) into a single medical population and all 37 novices into a single novice population. These two populations were fit independently. The (in)experienced medical participants were grouped together due to the practical limitations of hierarchal modeling; 9 and 10 participants in each sub-group respectively are insufficient to define a hierarchal population with the DDM. Given the high level of correlation between model parameters in this model, we utilize Differential Evolution Markov Chain Monte Carlo (DEMCMC) (Turner & Sederberg, 2012) to carry out this Bayesian estimation. Since the DDM does not have an analytically tractable closed form likelihood function, we utilize a recently developed approximation, the Probability Density Approximation (PDA) method (Holmes, 2015; Holmes & Trueblood, 2017; Turner & Sederberg, 2014), to approximate the likelihood of each parameter set sampled.

## Results and Discussion

We first examined average accuracy on the 60 practice trials preceding the main blast identification task to see how well participants learned to identify the images. For novice participants, the proportion of trials answered correctly in the practice block was 0.73 (SD = 0.09). We removed three participants with accuracy less than two standard deviations below the average because these participants were likely not engaged in the task. For the pathologists, the proportion of trials answered correctly in the practice block was 0.90 (SD = 0.08). One of the experienced pathologists was removed due to a computer error that affected data recording.

For the behavioral analyses, we used Bayesian statistics implemented using the open source software package JASP (Team, 2016). For each test, we report the Bayes factor (BF), which is the ratio quantifying the evidence in the data favoring one hypothesis relative to another (when comparing the alternative hypothesis to the null, we calculate $BF_{10}$ where the subscript '10' indicates evidence for the alternative '1' to the null '0'). While BFs are directly interpretable, labels for the strength of Bayes factors have been proposed. In particular, BF greater (less) than 1, 3 (1/3), 10 (1/10), 30 (1/30), and 100 (1/100) are considered 'Anecdotal', 'Moderate', 'Strong', 'Very Strong', and 'Extreme', evidence, respectively (Kass & Raftery, 1995).

First, we examined whether or not novice participants learned to generalize information about blast cells from training and practice to the main test trials. Because many of the images used in training and practice were also used in the main trials, it is possible that novices simply remembered specific images and their corresponding labels rather than learning general characteristics of blast versus non-blast cells. To examine this issue, we compared accuracy between the 'old' blast images (the 90 images used in training and practice) and the 'new' blast images (the additional 60 images not seen in training or practice) during the main trials. Overall, the accuracy on 'old' blast images during the main trials was 0.76 (SD = 0.14) and the accuracy on 'new' blast images was 0.75 (SD = 0.13). This difference was not statistically significant ($BF_{10}$ = 0.49; t(33) = 1.47, p = 0.152). Thus, we can conclude that participants did learn general characteristics of blast images during training and practice and were generalizing this information to new images during the main trials. Note that all 150 non-blast images were used in training and practice and thus this comparison is not possible for these images. However, we believe similar learning most likely occurred for the non-blast images rather than participants remembering individual images.

Next, we examined the hit and false alarm rates for the three groups of participants across all trials and conditions. We compared the hit rates for the three groups of participants using a Bayesian ANOVA. This analysis showed that the alternative model was strongly preferred to the null ($BF_{10}$ = 6032.67). In particular, the hit rate for both groups of pathologists was greater than the hit rate for novices ($BF_{10}$ = 221.5 for novices as compared to experienced pathologists; $BF_{10}$ = 262.3 for novices as compared to inexperienced pathologists), but there was no difference between the two groups of pathologists ($BF_{10}$ = 0.44). Next we compared the false alarm rates for the three groups using a Bayesian ANOVA and found that the alternative model was preferred to the null ($BF_{10}$ = 7.67). Specifically, experienced pathologists had a lower false alarm rate than novices ($BF_{10}$ = 38.63). However, there was no difference between the false alarm rate for novices and inexperienced pathologists ($BF_{10}$ = 0.43). There was also very little difference between the two groups of pathologists ($BF_{10}$ = 1.5).

*Signal Detection Theory Results*. We fit SDT to each of the speed, accuracy, and bias instruction conditions separately. We examined the best-fit values for the two key model parameters: discriminability and criterion. Figure 4 shows group-level posterior distributions (best fit parameter distributions) for these parameters in each of the three instruction conditions and for the three groups of participants and Table 1 lists the corresponding means. As shown in both Figure 4 and Table 1, experience leads to increased discriminability, but no change in criterion.

Next we analyzed differences in parameter values across instruction conditions by conducting Bayesian t-tests on the group-level posterior distributions and report the corresponding Bayesian p-values. For the two groups of pathologists, there was no significant difference in discriminability between speed and accuracy conditions ($p$ = 0.18 for experienced pathologists and $p$ = 0.15 for inexperienced pathologists). However, discriminability was significantly larger under accuracy instructions as compared to speed for the novice participants ($p$ = 0.03). There was no difference in the criterion for accuracy and speed instructions (all p-values were greater than 0.25)

For the bias condition, we fit trials where the cue was present and absent separately. Bayesian t-tests on the posterior distributions showed no difference in discriminability when the cue was present as compared to absent (all p-values were greater than 0.4). A Bayesian t-test on the posterior distributions showed the criterion was marginally lower when the cue was present as compared to absent for novices (p = .096). There was no difference in the criterion when the cue was present as compared to absent for the two pathologists groups (both p-values were greater than 0.195).

Table 1

*Means of the group-level posterior distributions for discriminability (top) and criterion (bottom) parameters from SDT for three groups of participants in the accuracy, speed, and bias conditions*

| Condition | Novice | Medical (< 4 rotations) | Medical (4+ rotations) |
|---|---|---|---|
| *Accuracy* | 1.45 | 2.22 | 2.73 |
| *Speed* | 1.22 | 1.86 | 2.50 |
| *Bias (cue present)* | 1.39 | 2.47 | 2.65 |
| *Bias (cue absent)* | 1.38 | 2.53 | 2.70 |
| *Accuracy* | -0.07 | -0.41 | -0.16 |
| *Speed* | -0.05 | -0.28 | -0.11 |
| *Bias (cue present)* | -0.19 | -0.31 | -0.18 |
| *Bias (cue absent)* | -0.06 | -0.51 | -0.08 |

In sum, the SDT analysis shows expertise influences discriminability and not criterion. However, we did not find any differences in the two key parameters across the instruction conditions (except for lower discriminability under speed instructions for novices). The lack of differences among instruction conditions is not surprising. SDT provides only a limited analysis of underlying cognitive processes in part because it does not take into account response times. Below, we analyze the data using the DDM, which takes into account both choice and response time data.

*Comparison of SDT parameters with visual ability (NOMT).* For all participants, the average proportion of correct responses on the NOMT was 0.73 (SD = 0.10). We compared participants' performance on the NOMT with the discriminability and criterion parameters from the SDT modeling. We calculated Bayesian Pearson correlations between NOMT accuracy and SDT parameters separately for accuracy, speed, and bias blocks since the model was fit separately to these conditions. The correlations are provided in Table 2. Overall, we found positive correlations between discriminability and NOMT accuracy. The correlations were the strongest for the speed and bias conditions (i.e., the Bayes factors for these conditions indicated 'moderate' evidence for the correlations). There was no evidence for correlations between NOMT accuracy and criterion. In sum, specific ability on the task (measured by discriminability) is positively related to domain general visual ability (measured by the NOMT).

Table 2
*Bayesian Pearson correlations between NOMT performance and discriminability and criterion parameters from SDT for the accuracy, speed, and bias conditions*

| Condition | Discriminability | Criterion |
|---|---|---|
| Accuracy | 0.23 ($BF_{10} = 0.64$) | 0.10 ($BF_{10} = 0.22$) |
| Speed | 0.31 ($BF_{10} = 1.85$) | 0.16 ($BF_{10} = 0.33$) |
| Bias (cue present) | 0.35 ($BF_{10} = 3.70$) | -0.07 ($BF_{10} = 0.19$) |
| Bias (cue absent) | 0.34 ($BF_{10} = 3.10$) | 0.04 ($BF_{10} = 0.18$) |

*Diffusion Decision Modeling Results*. Similar to SDT, we fit the DDM to each of the speed, accuracy, and bias instruction conditions separately. It is in principle possible to fit the totality of the data at once as is often done. Typically this is accomplished by fixing certain parameters (accumulation rates for example) to be the same across instruction conditions while others (threshold for example) are condition dependent. This however restricts up front the properties of the model that can vary between conditions. By fitting the three conditions separately, we allow maximal model flexibility, so that the data can determine what is the same or different across conditions.

In addition, we also fit the novice participants and pathologists separately. Note that for pathologists we only fit the hard trials in the accuracy condition. This is because the pathologists made almost no errors on the easy trials in the accuracy condition and the DDM has difficultly fitting data when choice proportions are near ceiling (i.e., perfect performance), since errors are required to inform some parameters. For the speed and bias conditions, the pathologists made a sufficient number of errors on the easy trials that we were able to include them in the fitting of these conditions.

To determine if the model was able to capture the data (which is necessary for it to be useful), we 1) extracted the mean parameters for fits to each of the conditions, 2) calculated the predicted choice proportion and mean response time (RT) for each condition, and 3) compared those predictions to choice proportions and mean RT's from data. Results (Figure 5) show that the model provides a good accounting of most aspects of the observed data. In each of these figure panels, the diagonal line represents the line of perfect agreement (prediction = observation), with results lying close to this line in most cases. Note there is more spread in the fits for the bias condition because this condition has half as many observations as the speed and accuracy conditions (due to the presence of cue / no cue trials). Thus increased noise in the data would be expected. Also, the model has some trouble accounting of the long response times, which is a common issue with DDM and other similar models since it is a model of speeded decision making.

Overall, the model fits the observed data well and thus we will further analyze the model results.

We next look at the best-fit values for the three key model parameters linked to behavioral characteristics of interest here (accumulation rate, threshold, bias). Figure 6 shows group-level posterior distributions (best fit parameter distributions) for these parameters in the speed and accuracy conditions. Analysis of the drift rate estimates shows a consistent pattern across conditions. For the novice participants, evidence accumulation rates are lower for more difficult images, regardless of instruction condition and difficulty. Note that we do not show the drift rates for the easy trials for the pathologists. These were not estimated for the accuracy condition and the drift rate posteriors for the speed and bias conditions were too broad (due to the small number of errors) to draw strong conclusion. Interestingly there is a significant difference between participants' ability to perceive the characteristics of blast and non-blast images respectively, as evidenced by the fact that $d_{NB} \neq d_B$. For novices, it appears that the characteristics of hard non-blast images are the most difficult to discern while the characteristics of easy non-blast are the simplest. In both novice participants (Figure 6 top panels) as well as pathologists (Figure 6 bottom panels), hard non-blast images were more difficult to discern than hard blast images.

Results additionally show that there is no detectable bias in the speed or accuracy conditions. That is, participants had no implicit preference for identifying cells as either blast or non-blast (i.e., posterior of the start-point bias includes 0). Comparison of the threshold parameters between the speed and accuracy conditions suggests that the speed instruction predominantly influences the threshold parameter. Thus, under speed instructions, it appears that both novice participants and pathologists become less cautious.

Figure 7 shows posterior distributions (best fit parameter distributions) for the bias condition. The introduction of a cue indicating a higher likelihood that the subsequent image is a blast does appear to introduce a small bias in both novices and pathologists. Specifically, the start-point parameter shifts toward the threshold in the presence of a cue suggesting that participants have a prior bias to respond 'blast' before seeing an image. In addition, the drift rates in the bias condition show a similar pattern to those in the speed and accuracy conditions.

*Comparison of DDM parameters with visual ability (NOMT).* Next, we compared participants' performance on the NOMT with measures of speed, accuracy, and bias derived from this modeling. To do so, we used Bayesian linear regression to predict NOMT performance using the best-fit parameters from the DDM for each individual (we included both novices and pathologists). We carried out the linear regression analyses separately for accuracy, speed, and bias blocks since the model was fit separately to these conditions. For the accuracy condition, there were 5 predictors ($t_{ND}$, $d_{BH}$, $d_{NBH}$, $a$, bias ($a-z/2$)) since $d_{BE}$ and $d_{NBE}$ where not estimated for the pathologists. For the speed condition, there were 7 predictors ($t_{ND}$, $d_{BE}$, $d_{BH}$, $d_{NBE}$, $d_{NBH}$, $a$, bias ($a-z/2$)). For the bias condition, there were 8 predictors since there were two different biases in the model (one for cued trials and one for uncued trials). We examined all possible combinations of predictors ($2^5=32$ models were fit for the accuracy condition, $2^7=128$ models were fit for the speed condition, and $2^8=256$ models were fit for the bias condition).

For the accuracy condition, no model was strongly preferred to the null model (for all models, $BF_{10} < 1.5$). For the speed condition, the preferred model was the one with only $d_{NBE}$ ($BF_{Model} = 10.41$ and $BF_{10} = 115.91$, $R^2 = 0.242$). In particular, participants with larger $d_{NBE}$ parameter values had better performance on the NOMT. For the bias condition, the preferred model was one with both non-blast drift rates ($d_{NBE}$, $d_{NBH}$) and the start-point bias parameter when the cue was absent ($BF_{Model} = 23.82$, $BF_{10} = 158.26$, $R^2 = 0.336$). Similar to the speed condition, participants with larger non-blast drift rates had better performance on the NOMT. In addition, better NOMT performance was associated with a smaller bias parameter value when the cue was absent. Overall, these results show that the primary cognitive DDM parameter that correlates with NOMT performance is the evidence accumulation rates on non-blast images. As compared to SDT, the DDM provides a more nuanced correlation between task specific ability and the NOMT, showing that the relationship is predominately driven by ability on non-blast images. These results suggest that the NOMT might have limited ability to identify individuals who make minimal detection errors (as this relationship seems to be confined to only non-blast images).

As a final note, we acknowledge that the difference in compensation between the pathologists and novices is a possible confound in our study. The pathologists were compensated with gift cards whereas the novices were compensated through course credit. While it is possible that our results were influenced by the difference in compensation, we feel that this effect was at most minor. In particular, the pathologists did not receive performance-based compensation. All pathologists received a gift card worth the same amount regardless of performance.

## Conclusions

In this study, we took a joint experimental and modeling approach to investigate the cognitive processes involved in cancer cell image detection in diagnostic pathology. To probe the differences between the underlying cognitive processes of novice and experts, we used SDT and DDM analysis to assess the influence of two common cognitive manipulations that are relevant in the clinical context – speed-accuracy tradeoff and prior expectations. Many medical image observers are facing increasing workloads due to the current and projected shortages of medical technologists and pathologists (Allen, 2002; Bennett, 2015; Lewin, 2016; Sullivan, 2016) along with desires to improve turnaround times and reduce costs. The aim to increase productivity can result in decreased screening times and ultimately a tradeoff between speed and accuracy. The increased reliance on automated and AI systems has the (counterintuitive) potential to compound the problem. Even though these systems can offset some the human workload, humans still play an integral role in diagnosis (at least for the near future). In the human-machine diagnostic team, it is often assumed that humans are doing less work per case and thus can increase the overall number of cases reviewed within a given day (e.g., FDA increased the workload for cytotechnologists from 100 slides per day to 200 slides per day if they are using the ThinPrep imaging system). However, such an increase in workload (even though it comes with the assistance of a machine) can potentially exacerbate the tradeoff between speed and accuracy.

Additionally, we assessed the influence of prior expectations on performance. In diagnostic pathology, images are often analyzed by medical technologists, residents, and / or automated basic recognition sorters before being seen by senior pathologists. Images that clearly lack abnormalities are rarely passed on to a senior expert. Thus, the mere presence of an image on an expert's desk is a cue, potentially setting expectations before the image is viewed. To examine the influence of prior expectations, we assessed how participants responded to the presence of a probabilistic cue. In addition, we also examined individual differences in decision-making by measuring domain general visual ability using the Novel Object Memory Test.

To assess the influence of these manipulations, we used two common modeling frameworks intended to extract cognitive parameters associated with task performance, Signal Detection Theory (SDT) and the Diffusion Decision Model (DDM). Each of these models was fit to participant data to assess how the parameters change in response to different instruction conditions (i.e., speed, accuracy, and bias conditions) as well as how parameter values relate to experience.

SDT shows a strong dependence of discriminability on expertise with increased expertise being associated with a higher degree of discriminability. There was no difference in the criterion parameter for different levels of experience. The SDT analysis also showed very little influence of instructions on parameters (though speed instructions appear to impact discriminability for novices). This finding is not surprising given the restricted nature of SDT, which only has two cognitive parameters to account for a wide array of potential effects and thus can lead to multiple effects being conflated. In particular, it has no mechanism for quantifying the effect of changes in cognitive strategies associated with response caution (which often occur under time pressure) or response biases. In addition, we found that NOMT performance was positively correlated with discriminability and not criterion.

DDM results paint a more detailed picture of the influence of the key manipulations (speed, accuracy, and bias) on cognitive processes. Results show that speed instructions lead to a significant reduction in caution in both novices and experts. We note that this finding is at odds with other literature suggesting that experts can become more accurate under speed instructions (Beilock et al., 2008; Beilock et al., 2004). In addition, estimates of the start-point bias parameter indicate that the presence of a probabilistic cue biases participants to respond 'blast' before viewing the image in both novices and pathologists. Finally, drift rate estimates show distinct differences between accumulation rates associated with blast and non-blast images. For difficult conditions, blast cells appear to be more discernable (higher associated drift rate) than non-blast in both novices and experts. In contrast, on easy conditions, non-blasts appear to be more discernable than blast cells for novices. We also examined the relationship between DDM parameters and NOMT accuracy. This analysis revealed that the primary DDM parameter that correlates with NOMT performance is the evidence accumulation rates on non-blast images. As compared to SDT, these results paint a more nuanced picture of the relationship between task specific ability and the NOMT, suggesting the NOMT might be limited in assessing individual differences in this task.

In aggregate, these results suggest the following conclusions. First, novices and experts have similar behavioral characteristics. While experts are clearly superior at the task (i.e., greater discriminability), both novices and experts respond to time pressure and external cues in similar ways and they both exhibit asymmetric responses to blast and non-blast stimuli. This suggests that while experiments with trained expert participants will always be the gold standard for research in this field, there is merit in working with novice participants, which are easier to recruit and allow for a wider array of studies. In addition, these results have important implications for training in this area. Clearly, expertise alone is not sufficient in altering the cognitive strategies and biases that are used when participants face time pressure and external cues. Second, our results show that individual differences in diagnostic decision-making are due in part to differences in visual ability (as measured by the NOMT), but these results are limited since the relationship is mainly driven by ability on non-blast images (as assessed by the DDM). Understanding individual differences is the first step in developing and improving individualized training. Future research could further explore the manipulations introduced here as well as the impact of individual differences in medical image decision-making.

# Declarations

**Ethics approval and consent to participate**

This study and all of its materials and consent documents were approved prior to the initiation of data collection by the Vanderbilt University Institutional Review Board (IRB # 161767).

**Consent for publication**

Not applicable

**Availability of data and material**

All de-identified data are available upon requests made to the corresponding author, jennifer.s.trueblood@vanderbilt.edu

**Competing interest**

The authors declare that they have no competing interests.

**Funding**

This work was supported by a Clinical and Translational Research Enhancement Award from the Department of Pathology, Microbiology, and Immunology, Vanderbilt University Medical Center. JST and WRH were supported by National Science Foundation grant SES-1556415.

**Authors' contributions**

All authors contributed to the study concept and design. The experimental program for the blast identification task was coded by MW and the experimental program for the NOMT was coded by WH, both under the supervision of JST and WRH. Testing and data collection was performed by MW, WH, ES, and JST. Data analyses were performed by JST. Computational modeling was performed by WRH and JST. WRH and JST drafted the manuscript, and all authors provided critical revisions. All authors approved the final version of the manuscript for submission.

**Acknowledgements**

The authors would like to thank Isabel Gauthier for advice on using the NOMT.

**Figures Captions**

Figure 1: Panels a-d) Sample images of blast and non-blast cells that were classified as easy and difficult. Panel a is an easy blast image, panel b is a hard blast image, panel c is an easy non-blast image, and panel d is a hard non-blast image. Panels e,f) Two sets of sample images from the Novel Object Memory Task (NOMT). Panel e shows the six Ziggerin targets and panel f shows the six Greeble targets.

Figure 2: Diffusion Decision Model Schematic. Evidence accumulates over time based on the stimulus present. Here the top / bottom boundaries indicate the levels of evidence needed to respond blast / non-blast respectively.

Figure 3: Signal Detection Theory Fit Results. Group-level posterior distributions for discriminability (top panels) and criterion (bottom panels) parameters for the three instruction conditions (speed and accuracy in the left panels and bias in the right panels) for the three groups of participants.

Figure 4: DDM Quality of Model Fit. Comparison of model predictions (vertical axes) and observed (horizontal axes) for response proportions (a-c for novices and g-i for pathologists) and mean response times (d-f for novices and j-l for pathologists) in the three instruction conditions. The solid diagonal line indicates perfect agreement where predictions and observations exactly coincide.

Figure 5: DDM Fit Results for speed and accuracy conditions. Posterior distributions for the threshold, bias, and evidence accumulation rate parameters (hyper mean parameters) for the speed and accuracy conditions (novices top panels and pathologists bottom panels). Threshold and start-point bias estimates are in the left panels with the accuracy (black) and speed (gray) conditions fit separately. Evidence accumulation rates are shown in the middle and right panels. Note that in the model, drift rates for blast and non-blast images are positive and negative respectively. Here we have presented $d_{BH}$, $d_{BE}$, $-d_{NBH}$, $-d_{NBE}$ for ease of comparison.

Figure 6: DDM Fit Results for bias conditions. Posterior distributions for the threshold, bias, and evidence accumulation rate parameters (hyper mean parameters) for the bias condition (novices top panels and pathologists bottom panels). Threshold and start-point bias estimates are in the left panels presenting estimates for the cued (red) and uncued (black) conditions separately. Evidence accumulation rates are shown in the right panel. Note that in the model, drift rates for blast and non-blast images are positive and negative respectively. Here we have presented $d_{BH}$, $d_{BE}$, $-d_{NBH}$, $-d_{NBE}$ for ease of comparison.

Figure 1

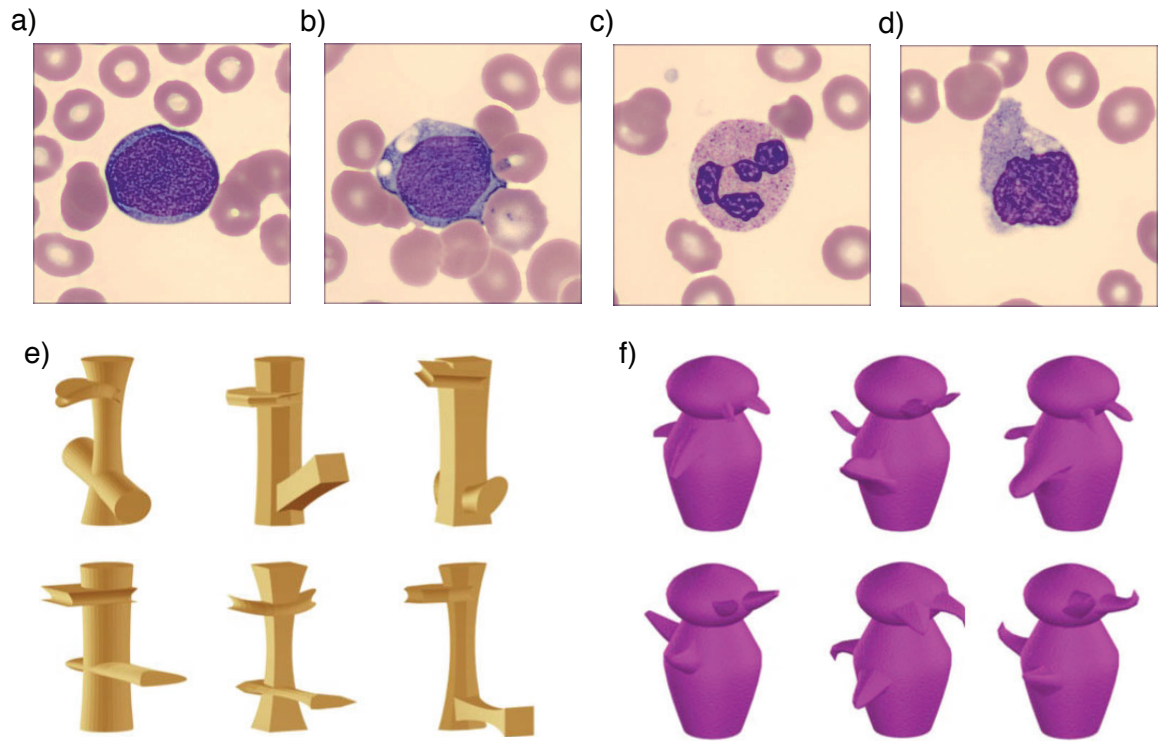

Figure 2

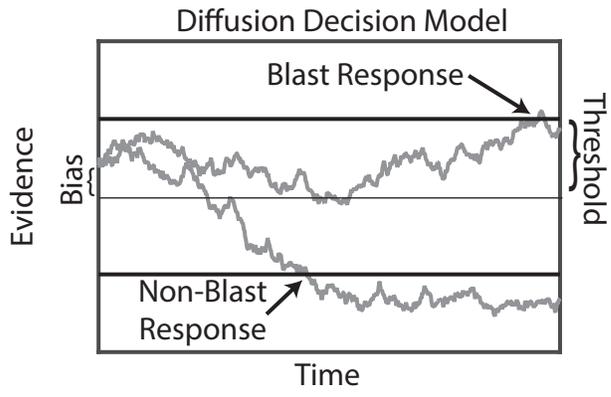

Figure 3

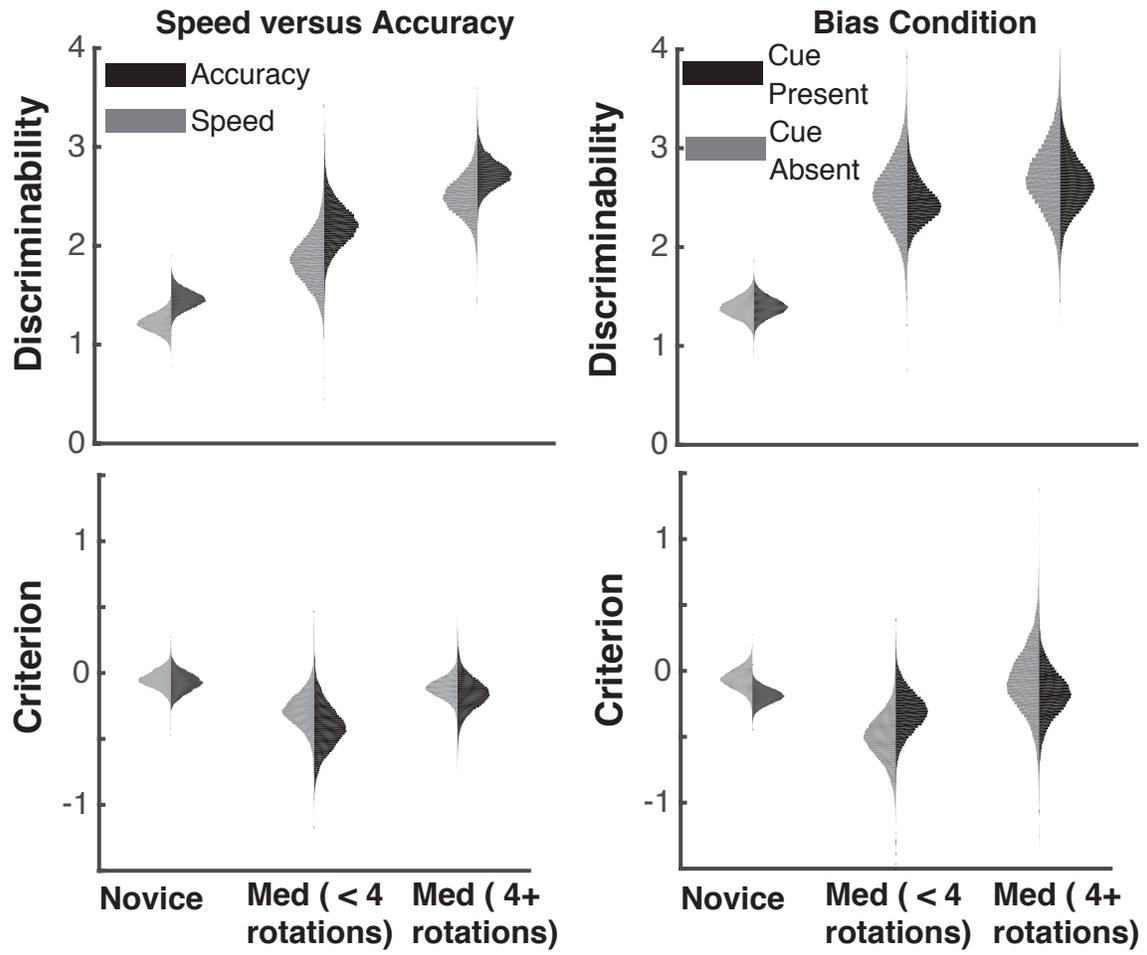

Figure 4

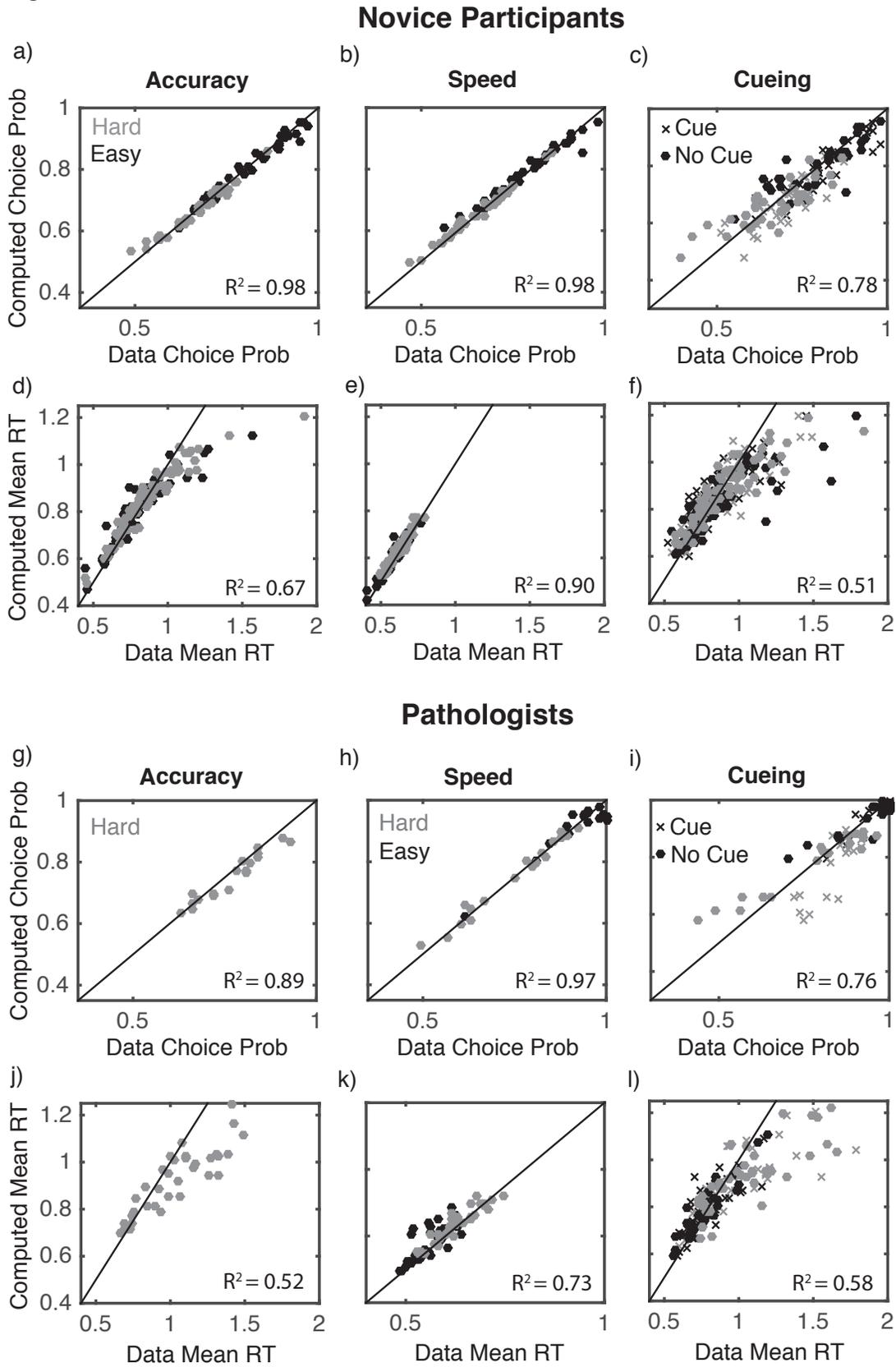

Figure 5

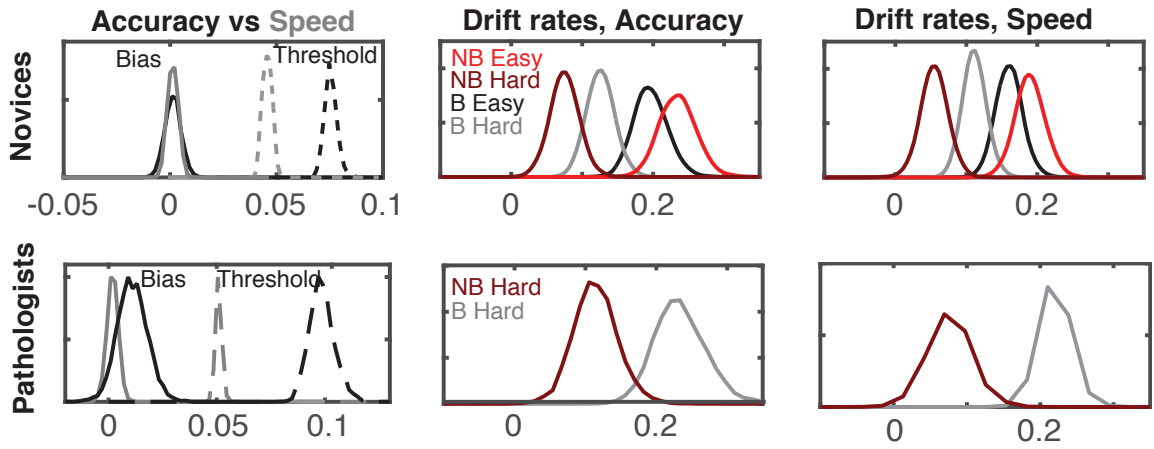

Figure 6

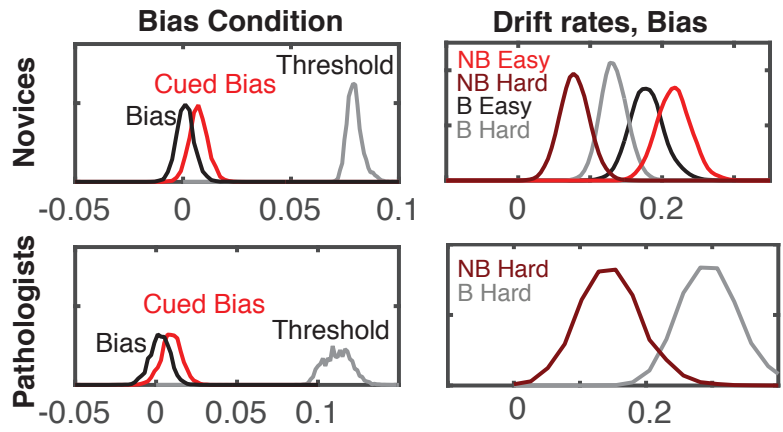

42